\documentclass[runningheads]{llncs}
\makeatletter
\newcommand{\@chapapp}{\relax}%
\makeatother

\usepackage[title,toc,titletoc,header]{appendix}
\usepackage{graphicx}
\usepackage{caption}
\usepackage{textcomp}
\usepackage{multirow}
\usepackage{amsmath}
\usepackage{subcaption}
\usepackage{pifont}
\usepackage{ctable}
\usepackage{booktabs,tabularx}
\usepackage{arydshln,leftidx,mathtools}

\usepackage{amssymb}
\begin{document}
\title{Grading Loss: A Fracture Grade-based Metric Loss for Vertebral Fracture Detection}

\titlerunning{\emph{Grading Loss} for vertebral fracture detection}
\author{Malek Husseini* \inst{1,2} 
\and Anjany Sekuboyina* \inst{1,2}
\and Maximilian Loeffler\inst{2} 
\and Fernando Navarro \inst{1,2}
\and Bjoern H. Menze\inst{1} 
\and Jan S. Kirschke\inst{2}}

%
\authorrunning{M. Husseini et al.}
%
\institute{ Department of Computer Science, Technical University of Munich, Germany \and
 Klinikum rechts der Isar, Technical University of Munich, Germany
\email{malek.husseini@tum.de} }


%
\maketitle              
\begin{abstract}
 Osteoporotic vertebral fractures have a severe impact on patients' overall well-being but are severely under-diagnosed. These fractures present themselves at various levels of severity measured using the Genant's grading scale. Insufficient annotated datasets, severe data-imbalance, and minor difference in appearances between fractured and healthy vertebrae make naive classification approaches result in poor discriminatory performance. Addressing this, we propose a representation learning-inspired approach for automated vertebral fracture detection, aimed at learning latent representations efficient for fracture detection. Building on state-of-art metric losses, we present a novel \emph{Grading Loss} for learning representations that respect Genant's fracture grading scheme. On a publicly available spine dataset, the proposed loss function achieves a fracture detection $F1$ score of 81.5\%, a 10\% increase over a naive classification baseline.
\end{abstract}

\renewcommand{\thefootnote}{\fnsymbol{footnote}}
\footnote[0]{* Shared first authors}

\section{Introduction}

Vertebral fractures are severely under-diagnosed. According to a 2013 study, 84\% of incidental vertebral fractures were not reported in CT \cite{5}. This is either due to the fractures being asymptomatic or to the symptoms wrongly being attributed to other factors. Osteoporotic vertebral fractures have critical consequences such as disability or increased mortality. Osteoporotic vertebral fractures cause pain and kyphosis in the short term, but are associated with an 8-fold higher mortality in the long term \cite{6}. Accentuating this is their high prevalence in older adult population (40\% by the age of 80 years), making a missed diagnosis critical. Therefore, there is a need for an automated and reproducible detection of vertebral fractures.

\subsubsection{Vertebral Fracture Detection} Automatic detection of vertebral fractures is relatively unexplored. Valetinitsch et al. \cite{9} propose the extraction of texture-based features such as histogram of gradients or local binary patterns from the trabecular of a segmented verebrae and classifying them using a random forest. From a deep learning perspective, Bar et al. \cite{10} employ a convolutional neural network for classifying sagittal patches from the vertebral column and aggregating the classification across patches using a recurrent neural network. Along similar lines, Tomita et al. \cite{11} work on thoraco-lumbar slices processed with a CNN and aggregated across slices using a long short-term memory (LSTM) network. However, unlike \cite{10}, the latter does not need any anatomy to be segmented to start the processing. Note that these approaches are \emph{ad hoc} implementations of CNNs working on large data samples and provide minimal insights into the workings of the network. Recently, Nicolaes et al. \cite{12} proposed a fully 3D approach for detecting vertebral fractures based on a voxel-level prediction regime, also providing a weak localization of the fracture. However, it being patch-based and predicting \emph{per voxel} limits its real-time applicability.

\begin{figure}[t]
    \centering
  
   \begin{subfigure}[b]{0.38\textwidth}
   \includegraphics[width=\textwidth]{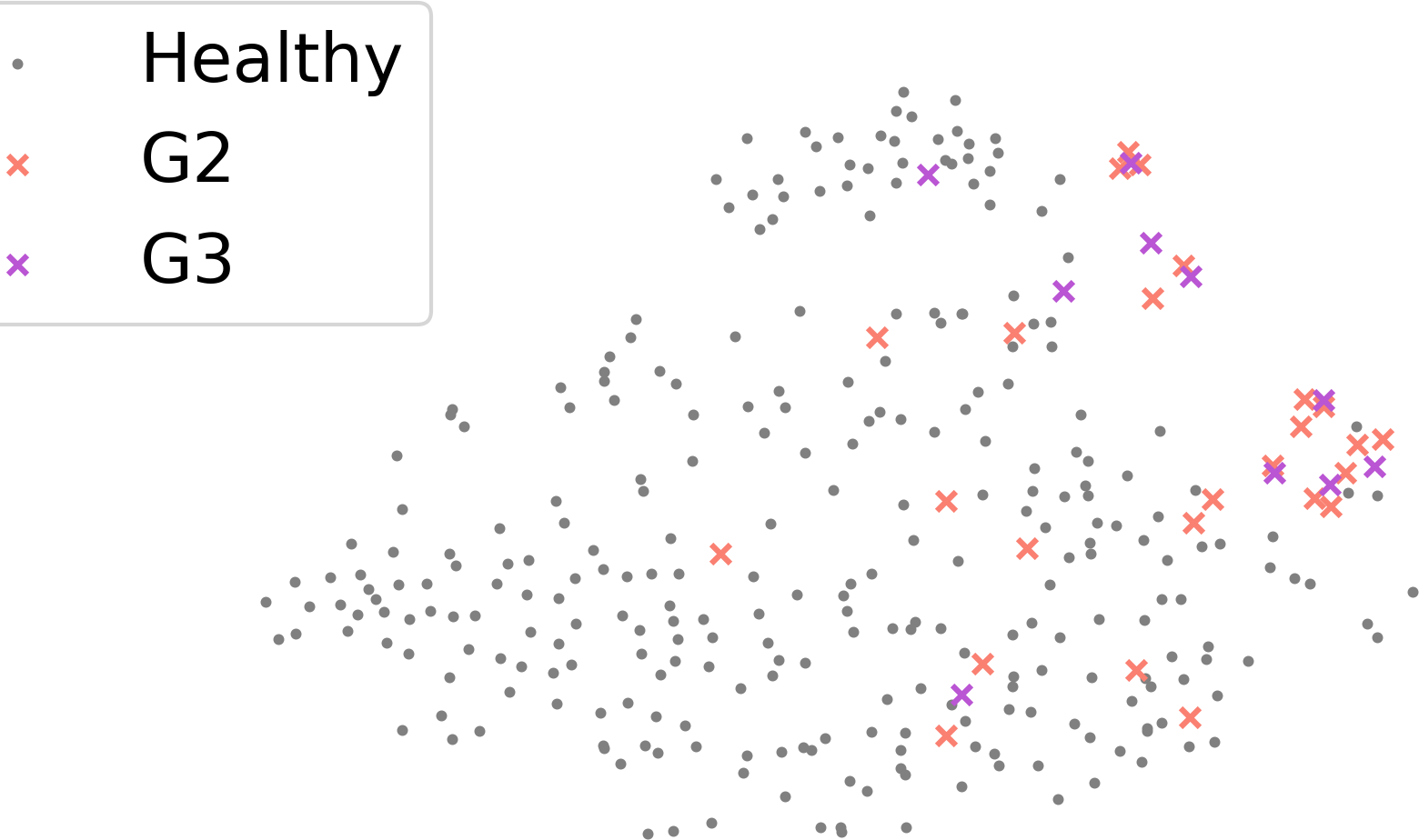}
        
        \caption{}
    \end{subfigure}
   ~
     \begin{subfigure}[b]{0.58\textwidth}
        \includegraphics[width=\textwidth]{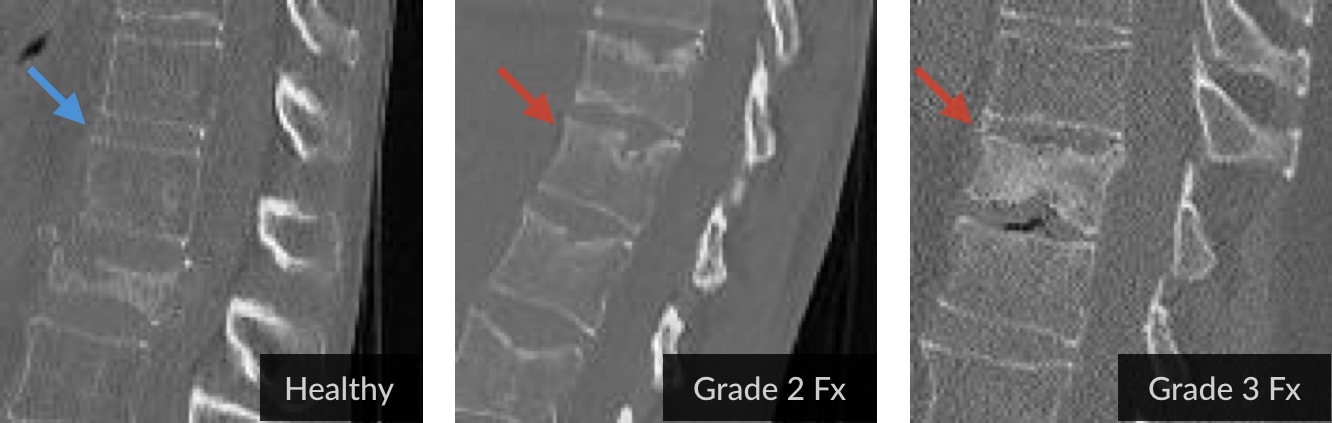}\vspace{0em}
        \caption{}
    \end{subfigure}
    \caption{\textbf{Illustrating fracture grades}: (a) TSNE visualisation of latent representations learnt by formulating fracture detection as a simple classification problem, resulting in poor separability. (b) An example selection of the three classes of vertebrae studied in this work, healthy, grade-2 fracture, and grade-3 fracture. }
	\label{fig:intro}
\end{figure}

We argue that formulating a vertebral fracture detection as a naive classification problem is sub-optimal, more so in case of limited and unbalanced data regimes. Fig.~\ref{fig:intro}a illustrates the TSNE representations of the latent features of one of the baselines in this work, viz. detecting vertebral fractures using a simple cross entropy loss using a convolutional neural network. Observe the resulting poor class-separation between healthy and fractured vertebrae. We attribute this to the wide variation in vertebral shapes: a healthy lumbar vertebra is `more different' from a healthy upper-thoracic vertebra than a fractured lumbar vertebra. Moreover, there exists a `gradation' among vertebral fractures, further obfuscating a clear shape-based separation (cf. Fig.~\ref{fig:intro}b).\\ 

\noindent
\emph{Genant's Vertebral Fracture Grading } The Current gold standard in grading vertebral fractures is a semi-quantitative method developed by Genant et al \cite{7}, according to which fractures are categorized into three grades (Grades 1, 2, and 3; cf. Fig \ref{fig:intro}b). This is based on the height-loss a vertebra undergoes compared to its healthy counterpart. 
A healthy vertebra is considered to be Grade 0. Grades 2 and 3 have proven clinical consequences, while this is unknown for Grade 1; as its small height reduction results in a high inter-rater uncertainty, Grade 1 fractures are excluded in this study. 

\subsubsection{Representation Learning} In this work, we aim to incorporate the gradual shape variations, courtesy of the fracture grades, into the training process of a classifier by explicitly adjusting the latent space. Deep learning models are believed to generate useful representations as a byproduct of the task they are trying to solve. However, this is not the case in low-data regimes as shown in Fig \ref{fig:intro}a. \emph{Representation learning} or \emph{metric learning} can be used to learn efficient latent representations in such scenarios. Siamese networks \cite{3} using contrastive loss and Face-Net \cite{2} with its triplet loss are examples of standard metric learning frameworks wherein representations of similar entities are clustered together while those of dissimilar ones are pushed apart.

\subsubsection{Contributions}
In this work, we attempt to solve vertebral fracture detection as a two-class, healthy \emph{vs.} fractured classification problem. \begin{itemize}
    \item Towards accurate classification, we pre-train the neural network using fracture-grade based representation learning. For this, we propose a novel \emph{loss function} termed \emph{grading loss}, which encourages the learnt representations to respect the gradation in the appearance of fractures.
    
    \item Accounting for the dependence of vertebral shapes on vertebral labels, we also propose a spine-region based pre-conditioning module.  
    
    \item We validate the proposed fracture detection regime on a publicly available VerSe dataset obtaining a classification $F1$ score of 82\%, outperforming naive classification as well as standard representation learning approaches.
    
\end{itemize}

\section{Methodology}
Given a collection of 2D vertebral patches, the objective of our work is to classify them into two classes, fractured and healthy. As vertebral shape depends on it label and the amount of variation in this shape due to a fracture depends on the fracture grade, we hypothesize that preceding the classification stage with fracture-grade and vertebral-label dependant pre-training results in an improved class separation. Consequently, this results in improved classification performance. We model these pre-training stages with inspiration from the field of representation learning.   

\subsection{Grading Loss}
\label{subsec:grading}
Metric learning aims to learn better data representations by working on a notion of `distance' in the latent space. It aims to cluster similar objects closer (by reducing the distance between them) while pushes dissimilar objects farther. This is done by optimizing loss functions, such as the contrastive \cite{3} and triplet losses \cite{2}. Note that these losses work on a notion of data similarity and dissimilarity. By design, they do not include a `ranking' within this similarity or dissimilarity. For example, a `ranking' is obvious in vertebral fractures, where a grade-2 and grade-3 vertebrae are \emph{fractured}, but the former is more similar to a vertebra from a \emph{healthy} class than the latter. Incorporating such 'ranking' criterion into the metric learning framework, we propose the \emph{grading loss}.

\begin{figure}[t]
\centering
\includegraphics[scale=0.8]{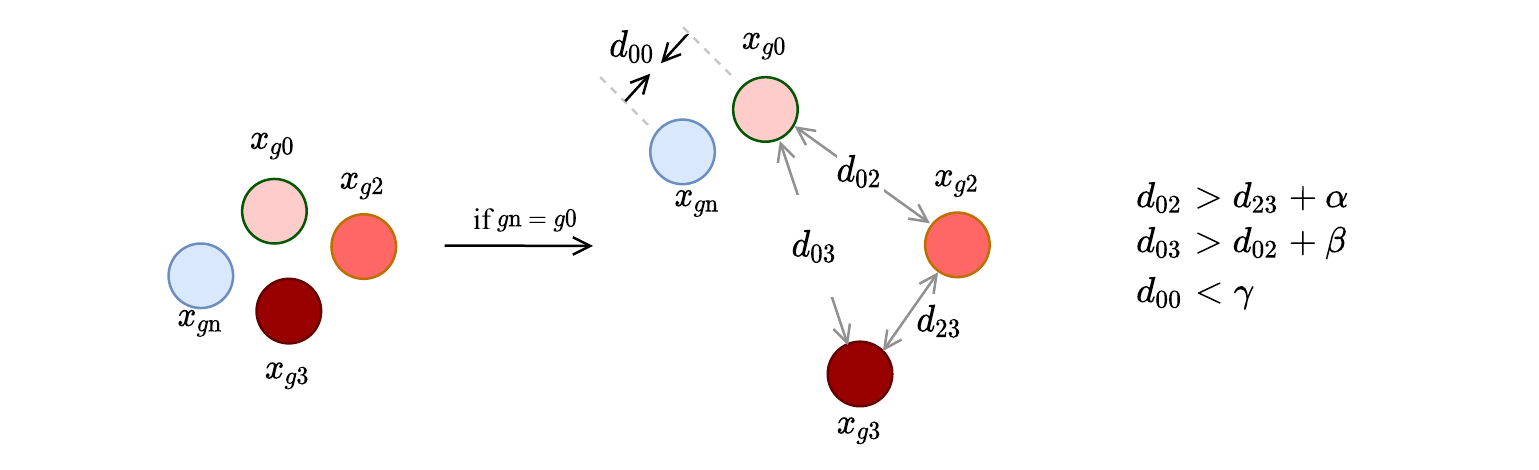}
\caption{Left: the arrangement of the three grades and the positive anchor during initialization. Right: $g3$ is drawn closer to $g2$ and both are pulled away from $g0$ under the constraint of $g0$ being closer to $g2$ than to $g3$. The positive anchor $gn$ works on clustering the similar classes together in the latent space. $gn$ belongs to the $g0$ sub-class in this figure.}\label{grading_loss}
\end{figure}

Assume a 2D vertebral patch, $x \in X$, is mapped to a representation $f(x)$ by a neural network $f$. The Euclidean distance between the representations of two examples $x^i$ and $x^j$ is denoted by $d(x^i, x^j) = ||f(x^i) - f(x^j)||_2^2$. We design the \emph{grading loss} as a quadruplet loss \cite{13} working with quadruplets denoted by $\{x_{g0}^i,x_{g2}^j, x_{g3}^k,x_{g\mathrm{n}}^l\}$, where $x_{g0}^i$ denotes a healthy vertebra sample, $x_{g2}^j$ and $x_{g3}^k$ denote samples of grade two and three, respectively, and $\mathrm{n}\in\{0,2,3\}$ can be randomly chosen as a healthy or a fractured example. Observe that the $x^i$, $x^j$, and $x^k$ form a static triplet, i.e. they are always sampled from fixed sub-classes of fracture grades. We incorporate a grading in the embedding space as follow: a grade-3 fracture is farther away from a healthy (grade-0) vertebra than a grade-2 fracture vertebra, and grade-2 is closer to grade-3 and it is to healthy. We can formulate these requirements as:
\begin{equation}
d(x_{g2}, x_{g3}) + \alpha < d(x_{g2}, x_{g0}) \textnormal{~and}
\label{eq:C1}
\end{equation}
\begin{equation}
d(x_{g0}, x_{g2}) + \beta < d(x_{g0}, x_{g3}),
\label{eq:C2}
\end{equation}
where $\alpha$ and $\beta$ are distance thresholds. Note that the Eq. \ref{eq:C1} uses $g2$ as a reference
sample and Eq. \ref{eq:C2} uses $g0$. Owing to the triangular inequality of distances, the restrictions on the distances from $g3$ to $g0$ and $g2$ are already satisfied by the above conditions. The above conditions can be achieved by optimizing the following loss terms:

\begin{equation}
\mathcal{L}_1 = \max (0 , d(x_{g2},x_{g3}) - d(x_{g2},x_{g0}) + \alpha) \textnormal{~and}
\label{eq:L1}
\end{equation}

\begin{equation}
\mathcal{L}_2 = \max (0 , d(x_{g0}, x_{g2}) - d(x_{g0},x_{g3}) + \beta).
\label{eq:L2}
\end{equation}

Observe that Eqs.~\ref{eq:L1} and \ref{eq:L2} structurally represent the triplet loss. However, observe that these do not work on similarities. They form \emph{separating} objectives between various fracture grades. Finally, a third, \emph{clustering} objective is incorporated by virtue of $x^l$ in the quadruplet. Recall that $x_{g\mathrm{n}}$ could belong to any of the three sub-classes. Based on the value of $\mathrm{n}$ in the sampled triplet, the clustering objective \emph{pulls} $x_{g\mathrm{n}}$ closer to its match in the static triplet. We demonstrate our \emph{grading loss} in Fig.\ref{grading_loss} with $x_{g\mathrm{n}}$ belonging to $g0$ sub-class and we refer to $x_{g\mathrm{n}}$ by the term positive anchor. The clustering objective can be represented as:
\begin{equation}
\mathcal{L}_3 = \max (0 , \gamma - d(x_{g\textrm{n}}^{\{i,j,k\}},x_{g\textrm{n}}^l))
\label{eq:L3}
\end{equation}

where $\textrm{n}\in\{0,2,3\}$ and $x^i$ and $x^j$ are a pair of samples from the same class. Assembling the loss terms together results in the proposed objective of \emph{grading loss}, $\mathcal{L}_G = \mathcal{L}_1 + \mathcal{L}_2 + \mathcal{L}_3$. In this work, the distance thresholds are chosen to be $\alpha > \beta > \gamma$. This is to ensure that $d(x_{g2}, x_{g0})$ is as large as possible while maintaining $d(x_{g3}, x_{g0}) > d(x_{g2}, x_{g0})$. That is, a higher separation between fractured and healthy classes is desirable compared to that between the two grades.

\subsection{Conditioning representations on vertebral indices}
\label{subsec:labels}
Considering the wide variation in shape from cervical to lumbar vertebrae, we claim that learning label-specific representations as a pre-training stage also improves fracture detection. Assuming the availability vertebral labels during the training process, we construct five categories of vertebra based on their shape similarity: $T1\sim T5,~T6\sim T9,~T10\sim T12,~L1\sim L4$, and $L5$. Treating this as a five-class problem, we can employ any standard metric loss for learning label-specific representations. Note that our \emph{grading loss} can also be extended for this case as there exists a `ranking' among the classes. Another application could be in brain tumors, where grades are also present (high grade and low grade gliomas). We leave the application of our loss to such scenarios for future work.

\subsection{Implementation} 

We perform fracture detection with the following network architecture \cite{1}, containing $5\times5$ filter kernels wherever applicable:\\

\noindent(conv32-bn-relu) \textrightarrow maxpool \textrightarrow (conv64-bn-relu)\textrightarrow maxpool
(conv128-bn-relu) \\ 
\textrightarrow maxpool \textrightarrow (conv256-bn-relu) \textrightarrow maxpool \textrightarrow (linear256-bn-lrelu)\\ 
\textrightarrow (linear128-bn-lrelu) \textrightarrow (linear64-bn-lrelu) \textrightarrow linear8\\

where `bn' and `lrelu' represent batch normalization and leaky Relu layers respectively. Recall that our network consists of a pre-training stage (for representation learning) and a training stage (for fracture detection). Furthermore, the pre-training stage consists of two sub-stages: vertebral-index-based representation learning and fracture-grade-based representation learning. Once pre-trained, the network is trained by optimizing a binary cross entropy loss over the fractured and healthy classes. For this, the last linear layer (linear8) is replaced with a two node linear layer (linear2) for the two classes. The network is implemented using the Pytorch library on an Nvidia GTX 1080 gpu. All losses are optimized using the Adam optimizer with a learning rate of $0.0001$. We set the hyper-parameters $\alpha, \beta$ and $\gamma$ to $1.5,1$ and $0.5$ respectively.

\section{Experiments}
In this section, we evaluate the contribution of the two main components proposed as part of our classification routine: first, the proposed grading loss' ability in in learning efficient representations, and second, our complete fracture detection routine.

\subsubsection{Dataset}
Recall that the proposed approach works at a vertebra level and utilizes vertebral labels. We utilize the publicly available VerSe \cite{8,15,16} dataset and its centroid annotations. As part of \cite{7}, its vertebra are annotated for fractures of three grades. We work with healthy, grade-2 and grade-3 fracture. We exclude the cervical vertebrae ($C1 \sim C7$) as vertebral fractures are extremely rare in this region. The dataset consists of 1283 vertebrae extracted from 157 scans, among which 1133 are healthy, 104 are $g2$ fractures and 46 are $g3$ fractures. The data is split into a training set containing 966 vertebrae and a test set with 312 vertebrae. The healthy:$g2$:$g3$ ratio in these sets is 851:79:36 and 282:25:10, respectively.

\subsubsection{Data Preperation}
Typically, a vertebra's mid-sagittal slice is a good indicator of a fractures. However, in cases where the vertebra presents itself in an atypical orientation, using the mid-sagittal slice is ineffective. Therefore, we utilize the vertebral centroids to extract 2D reformations of the vertebra along the mid-vertebral plane perpendicular to the vertebra's sagittal axis. Specifically, we construct a spline passing through the centroids and reformat the sagittal plane along which this spline passes. From this reformation, vertebral patches of size $112\times112$ pixels at $1\times1$ $mm$ resolution are extracted so that additional context is provided by the vertebra above and below the vertebra-of-interest (VOI). Our network consumes these patches. Additionally, a Gaussian around the centroid is passed as an additional channel for indicating the VOI.

\subsection{Experiments and Results}

We validate the proposed \emph{grading loss} in two stages: first, it is deployed as a stand-alone representation learning loss, where the separability of the learnt representations is tested (without any fracture-oriented training), and second, it is combined with a fracture classification module as a pre-training stage along with the proposed spine region-based representation learning component. The classification performances of various setups are compared using sensitivity ($SN$), specificity ($SP$), and $F1$ scores. We report the mean scores across fifteen randomly chosen folds of the dataset. Note that the proportion of healthy:$g2$:$g3$ is preserved through all the folds.

\begin{table}[t]
\centering
\setlength{\tabcolsep}{0.35em}
\small
\renewcommand{\arraystretch}{1.25}
\begin{tabular}{c c  c  c}
Setup &  $SN$ & $SP$ & $F1$  \\
\specialrule{.05em}{0em}{-.05em}
Contrastive & 54.2$\pm$5.3 & \textbf{95.6}$\pm$ \textbf{1.5}* & 57.4$\pm$4.2  \\
Triplet    & \textbf{71.7}$\pm$\textbf{5.1} & 90.2$\pm$ 2.0 & 57.3 $\pm$ 3.3 \\ 
\textbf{Grading}    & 67.1$\pm$0.2 & \textbf{95.3}$\pm$ \textbf{1.6}* & \textbf{65.2} $\pm$ \textbf{5.1} \\ 
\specialrule{.1em}{0em}{1em}
\end{tabular}
\caption{Evaluating learnt representations (\textbf{representation learn \textrightarrow fracture train}): Performance comparison of various losses for learning fracture-specific representations. * indicates statistical insignificance ($p$-value=0.44).}
\label{table:representation}
\end{table}

\subsubsection{\emph{Grading loss} results in better representations}
\begin{figure}[t]
    \centering
  
   \begin{subfigure}[b]{0.3\textwidth}
        \includegraphics[width=\textwidth]{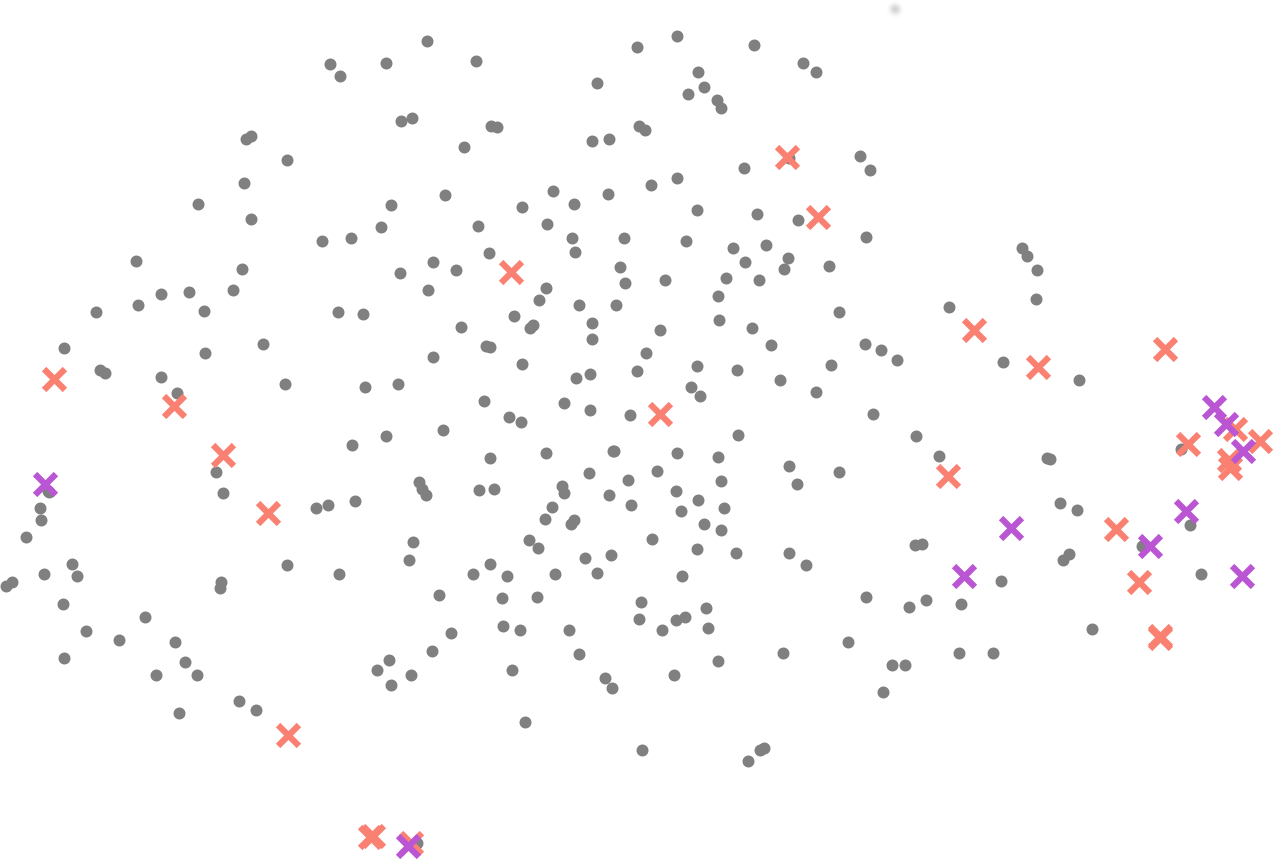}
        \caption{\small{Contrastive Loss}}
    \end{subfigure}
   ~
     \begin{subfigure}[b]{0.3\textwidth}
        \includegraphics[width=\textwidth]{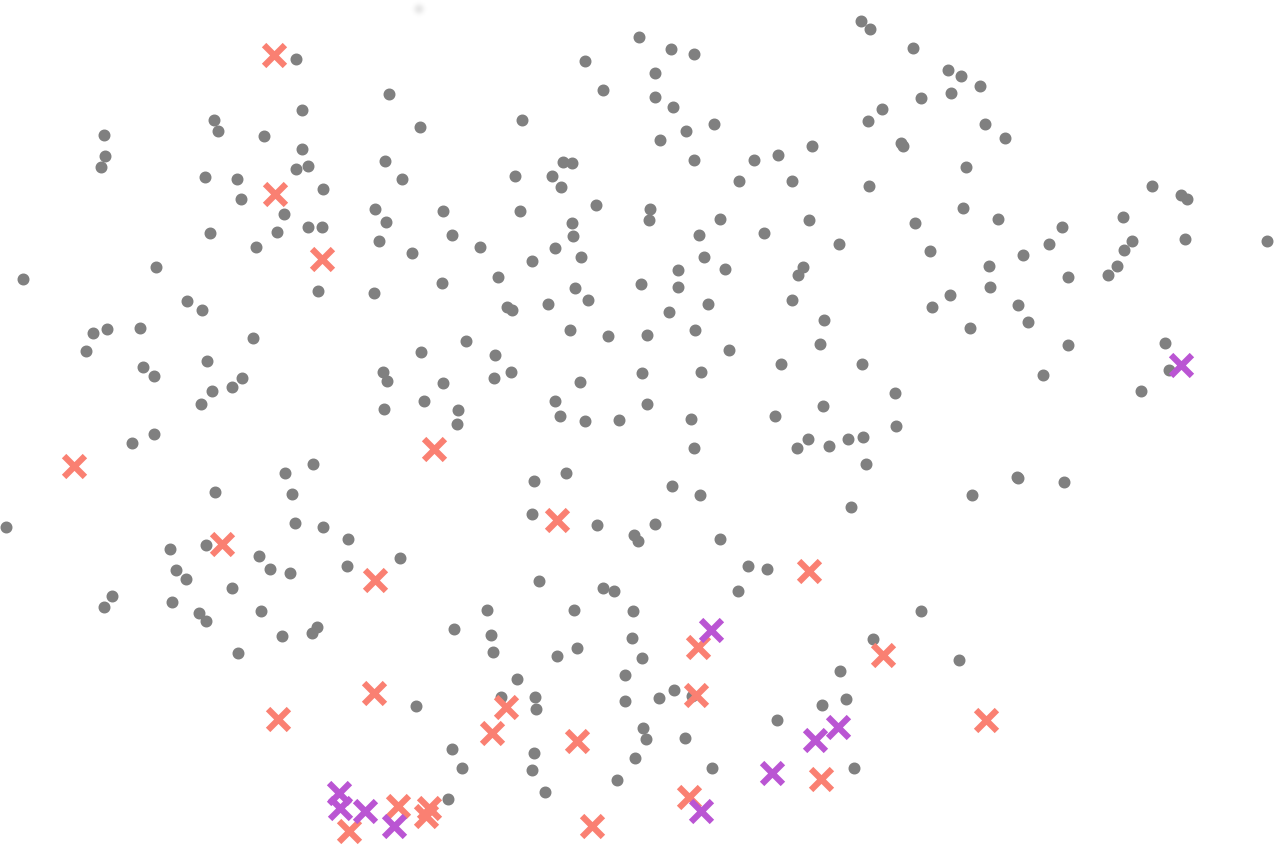}
        \caption{\small{Triplet Loss}}
    \end{subfigure}
   ~
    \begin{subfigure}[b]{0.3\textwidth}
        \includegraphics[width=\textwidth]{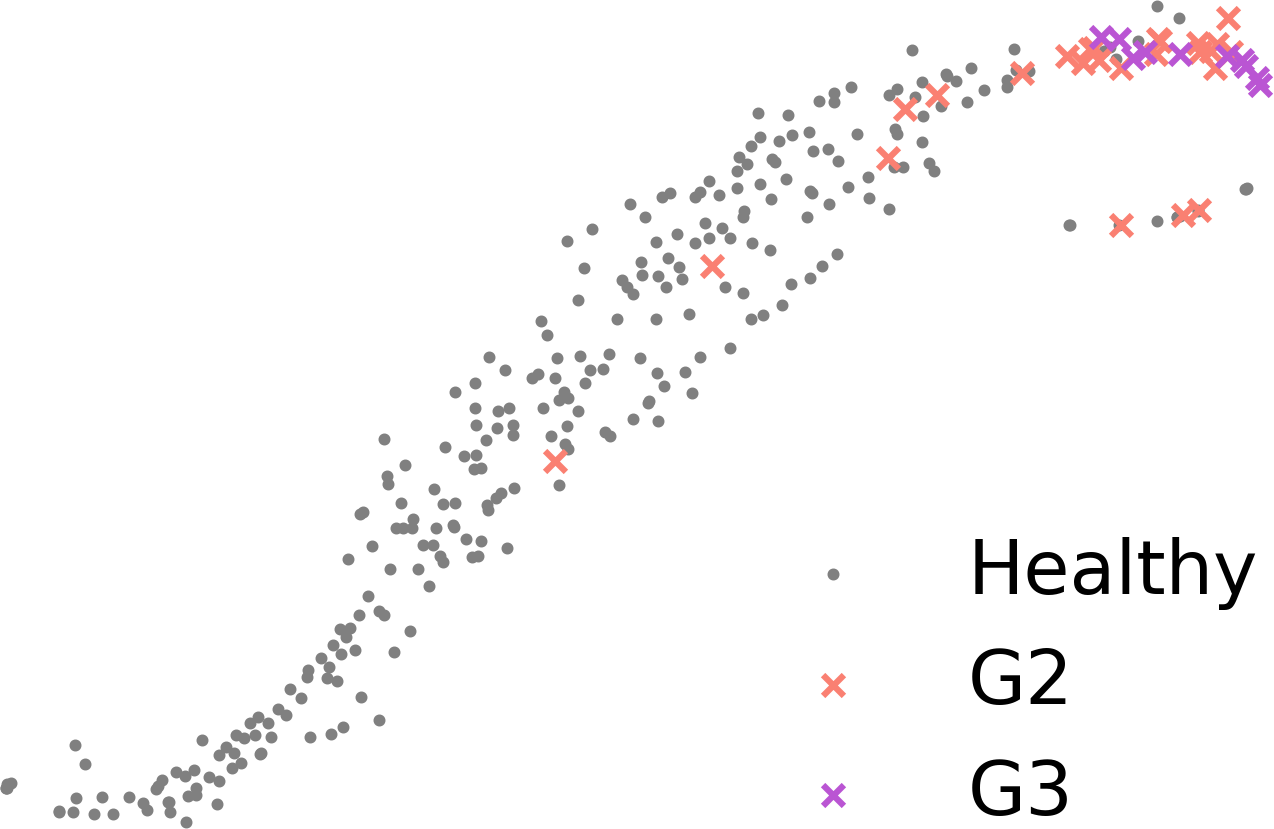}
        \caption{\small{\emph{Grading loss}}}
    \end{subfigure}
    \caption{TSNE visualisation of the representations learnt by various metric learning losses, without explicit classification-specific training. Proposed \emph{grading loss} obtains more separability between healthy and fractured classes.}
	\label{fig:representation}
\end{figure}

In this experiment, we validate the effectiveness of the proposed \emph{grading loss} at learning efficient representations for fracture detection.  We compare our loss with the two standard metric-learning losses: contrastive and triplet losses. Specifically, the neural network is optimized on the training set using each of the metric losses. Once trained, it is used to obtained the latent representations of the training samples on which a support vector machine (SVM) with a linear kernel is learnt. The more linearly separable the representations are, better the learnt SVM performs on the test set's latent representation. Table~\ref{table:representation} reports the classification performance of this SVM on the test set representations. Observe that the \emph{grading loss} readily offer better `linear' separability ($\sim 8\%$ increase in $F1$ score) of the fracture vs. healthy classes compared to the contrastive and triplet loss. The TSNE visualsations of these representations (cf. Fig.~\ref{fig:representation}) illustrate this clustering characteristic of the \emph{grading loss}. 

\begin{table}[t]
\centering
\setlength{\tabcolsep}{0.5em}
\small
\renewcommand{\arraystretch}{1.25}
\begin{tabular}{c c c c c  c}
Label Pre-train & Rep. Learn. & Frac. train & $SN$ & $SP$ & $F1$ \\
\specialrule{.05em}{0em}{-.05em}
\ding{56} & \ding{56} & \checkmark & 71.0$\pm$6.4 & 96.4$\pm$1.7 & 71.2 $\pm$ 4.4 \\
\checkmark & \ding{56} & \checkmark & \textbf{72.4} $\pm$ \textbf{7.7} & \textbf{97.8} $\pm$ \textbf{1.0} & \textbf{75.9} $\pm$ \textbf{5.5} \\
\hdashline
\ding{56} & Contrastive & \checkmark & 73.7 $\pm$ 7.7 & 98.0 $\pm$ 1.1 & 77.6 $\pm$ 2.6 \\
\ding{56} & Triplet & \checkmark & 73.7$\pm$6.3 & 96.4$\pm$1.4 & 73.6$\pm$ 2.4 \\
\ding{56} & \textbf{Grading} & \checkmark & \textbf{76.0} $\pm$ \textbf{5.8} & \textbf{97.8} $\pm$ \textbf{1.1} & \textbf{78.6} $\pm$ \textbf{4.5}\\
\hdashline
\checkmark & Contrastive & \checkmark & 74.3 $\pm$ 4.7 & 98.0 $\pm$ 0.6 & 78.2 $\pm$ 2.1 \\
\checkmark & Triplet & \checkmark & 75.7 $\pm$ 6.4 & 97.6 $\pm$ 0.8 & 77.5 $\pm$ 2.9 \\
\checkmark & \textbf{Grading} & \checkmark & \textbf{76.9}$\pm$ \textbf{5.8} & \textbf{98.5}$\pm$ \textbf{0.95} & \textbf{81.5} $\pm$ \textbf{3.8} \\
\specialrule{.1em}{0em}{1em}
\end{tabular}
\caption{Validating the proposed fracture detection regime (\textbf{label pre-train~\textrightarrow~representation learn\textrightarrow~fracture train}): Comparison of the proposed training routine based on \emph{grading loss} with naive classification as well as with other representation-learning-augmented classifications.}
\label{table:frac_routine}
\end{table}

\subsubsection{Proposed fracture detection regime} 
 Our complete fracture detection pipeline consists of three stages: two pre-training stages followed by the main classification stage. The first pre-training stage includes optimizing a contrastive loss over the five regions of spine described in Sec. \ref{subsec:labels}. Experiments justifying the choice of contrastive loss for this stage are presented in the supplement. Following this, the network goes through the second pre-training stage where our \emph{grading loss} is minimized. Finally, the network is optimized for fracture detection using cross entropy loss. We represent the proposed pipeline as \textbf{label pre-train~\textrightarrow~representation learn\textrightarrow~fracture train}). Table~\ref{table:frac_routine} reports an ablative test of the proposed routine. We test the contribution of label-based pre-training and that of the proposed grading loss-based representation learning is evaluated. Compared with a baseline network trained end-to-end for fracture detection, pre-training with vertebral labels offers a 5\% improvement in $F1$ score. On a different note, tuning the representations with fracture grades using \emph{grading loss} also improves the classification performance, providing about 7\% $F1$ score improvement over simple classification. These validate the effectiveness of the two pre-training stages incorporated in our routine. Finally, testing all the stages of the proposed pipeline, we observe that our combination results in the highest performance across the three metrics with an overall $F1$ score of 81\%. 

\section{Conclusion}

We conclude that in case of low-data regimes with severe data imbalance, augmenting classification with representation learning-based pre-training helps. Compared to conventional metric losses which work on similarity or dissimilarity of examples, the proposed \emph{grading loss} which incorporating a `ranking' within the classes provides a superior performance. Going a step further, incorporating the vertebral label information using similar techniques of representation learning further improves fracture detection. The proposed fracture routine achieves an $F1$ score of 81.5\%, an improvement of over 10\% over naive classification baseline. In future work, we will extend this study to incorporate grade-1 fractures as well as a 3D context, thus making our approach more robust to severely deformed spines.

\subsection*{Acknowledgements}
This work is supported by DIFUTURE, funded by the German Federal Ministry of Education and Research under (01ZZ1603[A-D]) and (01ZZ1804[A-I]).

\begin{appendices}
\renewcommand{\thesection}{\appendixname~A}

\section{Supplementary Material: Label pre-train}

\begin{table}[h]
\centering
\setlength{\tabcolsep}{0.35em}
\small
\renewcommand{\arraystretch}{1.25}
\begin{tabular}{c c  c  c}
Setup &  $SN$ & $SP$ & $F1$  \\
\specialrule{.05em}{0em}{-.05em}

 Cross-Entropy & 72.1$\pm$5.4 & 96.6$\pm$1.2 & 72.4 $\pm$ 5.7 \\
 Contrastive & \textbf{72.4}$\pm$\textbf{7.7} & \textbf{97.8} $\pm$ \textbf{1.0} & \textbf{75.9} $\pm$ \textbf{5.5} \\
 Triplet & 68.9$\pm$6.9 & 97.0$\pm$1.3 & 71.4$\pm$3.9 \\
 \hline
 
\end{tabular}
\bigskip
\label{table:3}
\caption{Contrastive loss was used in the \textbf{label pre-train} stage in our work. Validating its need and effectiveness, we compare it against two other pre-training approaches: (2) Classification-based: pre-training to identify the five spine region a given vertebra belongs to, followed by training for fracture-classification. (2) Representation learning-based:  pre-training to learn representations towards identifying the spine region  using triplet loss,  followed by training for fracture-classification. Our contrastive-loss based pre-training  results in an F1-score of 75.9, better than pre-training using cross-entropy loss or triplet loss.}
\end{table}

\end{appendices}

\end{document}